\documentclass{emulateapj}
\usepackage{epsfig}
\usepackage{amsmath}
\usepackage{natbib}
\usepackage{apjfonts}
\def\kmsmpc{{\,\rm km\,s^{-1}Mpc^{-1}}} 
 \def\kms{{\,\rm km\,s^{-1}}}

\begin{document}

\title{The bimodal galaxy color distribution: dependence on luminosity  and environment}
\shortauthors{Balogh et al.}
\shorttitle{The bimodal galaxy color distribution}
\author{Michael L. Balogh,\altaffilmark{1,2} Ivan
  K. Baldry,\altaffilmark{3} Robert Nichol,\altaffilmark{4} Chris Miller,\altaffilmark{4}
  Richard  Bower\altaffilmark{2} \& Karl Glazebrook\altaffilmark{3}\\}

\altaffiltext{1}{Department of Physics, University of Waterloo,  Waterloo, Canada N2L 3G1 E-mail: mbalogh@uwaterloo.ca}
\altaffiltext{2}{Department of Physics, University of Durham, South
  Road, Durham DH1 3LE, UK Email: r.g.bower@durham.ac.uk}
\altaffiltext{3}{Department of Physics and Astronomy, Johns Hopkins
  University, 3400 North Charles Street, Baltimore, MD 21218-2686 USA.
Email: baldry,kgb@pha.jhu.edu}
\altaffiltext{4}{Department of Physics, Carnegie Mellon University,
  5000 Forbes Avenue, Pittsburgh, PA 15213 Email: nichol,chrism@cmu.edu}

\begin{abstract}
We analyse the $u-r$ color distribution of 24346 galaxies with
$M_r\leq-18$ and $z<0.08$, drawn from the Sloan Digital Sky Survey
first data release, as a function of luminosity and environment.  The
color distribution is well fit with  two Gaussian
distributions, which we use to divide the sample into a blue and
red population.  At fixed luminosity, the mean color of the blue (red) distribution
is nearly independent of environment, with a weakly significant ($\sim3\sigma$) detection
of a trend for colors to become redder by $0.1$--$0.14$ ($0.03$--$0.06$) mag with a factor $\sim 100$ increase in
local density, as characterised by the
surface density of galaxies within a $\pm1000$ km~s$^{-1}$ redshift
slice.  In contrast, at fixed luminosity the fraction of galaxies in the red
distribution is a strong function of local density, increasing from
$\sim 10$--$30$ per cent of the population in the lowest density
environments, to $\sim 70$ per cent at the highest densities.  The strength of this trend is similar
for both the brightest ($-23<M_r<-22$) and faintest ($-19<M_r<-18$)
galaxies in our sample.  
The fraction of red galaxies within the virialised regions of clusters
shows no significant dependence on velocity dispersion.  
Even at the lowest densities explored, a substantial population of red galaxies
exists, which might be fossil groups. 
We propose that most star-forming galaxies today
evolve at a rate that is determined primarily by their intrinsic properties, and
independent of their environment.  Any environmentally triggered
transformations from blue to red colors must either occur
on a short timescale, or preferentially at high redshift, 
to preserve the simple Gaussian nature of the color distribution.  
The mechanism must be effective for both bright and faint galaxies.
\end{abstract}
\keywords{galaxies: clusters --- galaxies: evolution }
\maketitle

\section{Introduction}
The local galaxy population is known to consist broadly of
two types, identifiable for example by their morphological properties, and are termed late-type
(spiral) and early-type (E/S0) galaxies.  Recently, this
division has been quantified in large datasets, in the related
quantities of broadband color
\citep{Strateva01_short,Blanton03BB,Kauffmann-SDSS1_short}
and star formation rate \citep[SFR,][hereafter Paper~II]{Brinchmann03,2dfsdss}.  In particular, the
color distribution at fixed luminosity is surprisingly well modelled
by only two Gaussian distributions \citep[][hereafter Paper~I]{Baldry03}.  The mean and
variance of these two distributions are  strong functions of
luminosity, or stellar mass
\citep[see also][]{Bernardi03_short,Blanton03BB,Hogg03}; a similar trend is seen
out to $z\sim 1$ \citep{BellGems2}.  

\citet{Dressler} was first to show that the fraction of late--type
galaxies depends strongly on local galaxy environment, and related
trends in SFR have also been observed
\citep[e.g. Paper~II,][]{Miller_agn}.   If these trends are due to an
environmentally--induced transformation, we can look for clues to
its nature in the change of the properties of
galaxies within each type as a function of environment.
For example, if the transformation is due to
a decreasing SFR in late--type (blue) galaxies due to
interactions with neighbouring galaxies, then
their color distribution should be skewed toward redder colors in denser environments.  

Many studies of the fundamental plane
have shown that the properties of early--type galaxies are
nearly independent of environment
\citep[e.g.][]{Dressler87,BernardiIII_short}. The properties
of the blue population as a function of environment have been studied
in less detail;  early suggestions that their colors
do not depend on environment were seen in the small sample of \citet{LTC}.
This result has been supported by analysis of the Tully--Fisher
relation in clusters, groups and the field, using samples of several
hundred galaxies.  In particular, the scatter in this
relation is related to galaxy color, and it has been shown to be insensitive to environment
\citep[e.g.][]{GMM,BGMM}.   

Using galaxy colors to describe the galaxy population, rather than
morphological types, has the advantage 
that they are easily quantifiable, the measurements are robustly
reproducible, and models exist which allow us to directly relate them
to star formation histories with minimal assumptions \citep[e.g.][]{BC03}.
In particular, in Paper~I we have fit the color distributions of galaxies
selected from the Sloan Digital Sky Survey
\citep[SDSS,][]{SDSS_tech_short} with a two-component Gaussian model,
as a function of luminosity.  This technique has an
advantage relative to using strict color cuts to define separate
populations \citep[e.g.][]{BO84,KB01}, as it accomodates photometric 
and astrophysical uncertainties, such as aperture effects
\citep{Bernardi_CM_short}, dust reddening and stochastic variations in
star-formation history.  Furthermore, no a priori assumption about
how to classify the population is required.   

In Paper~I we used this novel procedure to show that
the luminosity function of red galaxies is consistent with a model in
which they are built up from mergers of blue galaxies.
In this {\it Letter}, we develop this investigation by exploring how
the model fits depend on local environment.

\begin{figure*}
    \center\leavevmode\epsfxsize=16cm\epsfbox{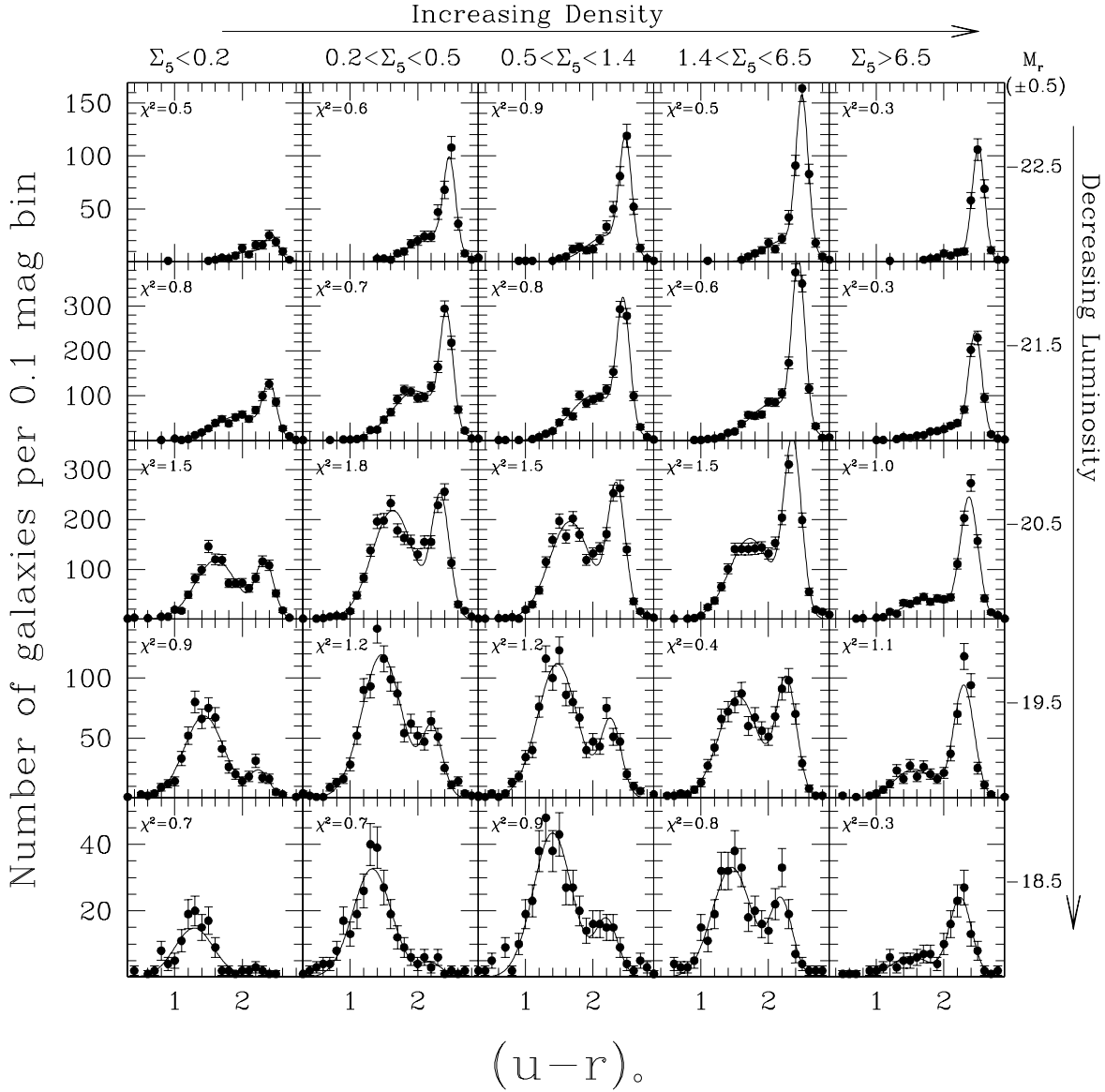}
    \caption{The {\it solid points} in each panel show the galaxy color
      distribution for the indicated 1 magnitude range of
      luminosity (right axis) and the range of local projected density, in
      Mpc$^{-2}$, shown on the top axis.  1-$\sigma$ error bars
      are given by $\sqrt{N+2}$, where $N$ is the number of galaxies
      in each bin.  
      The {\it solid line} is a double Gaussian model, with the
      dispersion of each distribution a function of luminosity only.  The reduced $\chi^2$ value of the fit is shown in
      each panel.
\label{fig-histograms}}
\end{figure*}
\section{Data}
To allow us to reliably measure local galaxy densities, we select a
sample of galaxies from the first data 
release of the SDSS \citep{DR1}, in the redshift range
$z<0.08$, with luminosities $M_r<-18$, assuming a $\Lambda$CDM
cosmology with $\Omega_m=0.3$, $\Lambda=0.7$ and $H_\circ=70 \kmsmpc$.  
Galaxy colors are measured from
model magnitudes \citep{EDR},
corrected for Galactic extinction using the dust maps of \citet{SFD}
and k-corrected to $z=0$ using the \citet{Blanton03} model. 
Projected local densities, $\Sigma_5$, are computed from the distance to the fifth
nearest neighbour that is brighter than $M_r=-20$ and within a redshift 
slice $\pm 1000\kms$ of each galaxy, as described in Paper~II.  
Applying this
bright limit to the density calculation gives us a uniform density estimate
that is applicable to our magnitude
limited sample over the full redshift range.
To ensure robust measurements, we only consider galaxies
sufficiently far from the survey boundary that the density estimate is
unbiased, which limits our sample to 24346 galaxies.  As a measurement of
the environment on large scales, we use the C4 
catalogue, based on the SDSS (Gomez et al.\ 2003, Miller et
al.\ in prep), to identify galaxies that lie in groups or clusters.
Cluster members are defined to be those within the 
virial radius and within 1000$\kms$ of the average cluster redshift.

\section{Results}
Figure~\ref{fig-histograms} shows the $u-r$ color distribution of galaxies
in bins of local density and luminosity.  The population is divided
into five density bins with the three middle
bins having an equal number of galaxies and the least and most dense having half
as many, to sample the extremes of the distribution.  The
median density of galaxies in the sample is $\Sigma_5\sim 0.8$
Mpc$^{-2}$; thus, the lowest density bin (median density $\sim 0.1$
Mpc$^{-2}$) is underdense by a factor
$\sim 8$, while the densest bin corresponds to the typical density
found in cluster cores.
Following Paper~I, we model the distributions
with two Gaussians using a Levenberg-Marquardt
algorithm.  The mean and amplitude of each distribution are varied
as a function of luminosity and local density, while the
dispersions are constrained to be a function of luminosity only.
The optimal parameters are found by minimizing the 
$\chi^2$ fit to the data in Figure~\ref{fig-histograms}; these
best-fit models are shown as the solid lines.

The most striking result of Figure~\ref{fig-histograms} is that the
double--Gaussian model provides an acceptably good fit to all the distributions, with
reduced $\chi^2\sim1$. Specifically, there is no convincing evidence that the
shape of the blue distribution is significantly distorted in dense environments
as might be expected if these galaxies were slowly being transformed
into the late-type (red) population.

\begin{figure}
    \leavevmode\epsfxsize=8cm\epsfbox{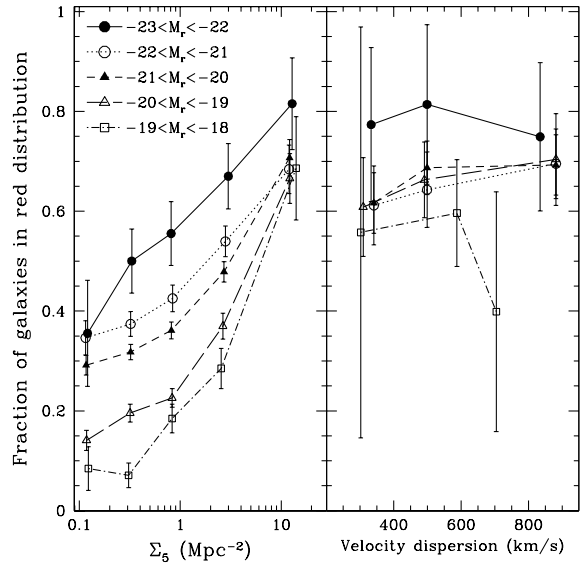}
    \caption{The fraction of galaxies in the red distribution,
      in bins of local density ({\it left panel}) or cluster velocity dispersion
      ({\it right panel}), based on the Gaussian fitting of
      Fig.~\ref{fig-histograms}.  Five lines are shown, corresponding
      to five  different luminosity ranges, as indicated.  Points are
      shown with $1-\sigma$ error bars at the median density or
      velocity dispersion of each bin.  Each point represents the
      fraction of red galaxies only among galaxies with that luminosity
      and environment and, therefore, they do not need to add to 100 per cent.
\label{fig-fr}}
\end{figure}

The dominant change in the galaxy population as a function of
environment is in the relative number of galaxies in each distribution,
as seen in earlier morphological  \citep[e.g.][]{Dressler} and emission--line
(e.g. Paper~II) studies.  To show this explicitly, we divide the sample
into bins of luminosity and environment (either local density or
group velocity dispersion) and compute the
relative abundance of red galaxies within each bin; the results are
shown in Figure~\ref{fig-fr}.  
There is a strong and continuous dependence on local density, with the
fraction of galaxies in the red distribution at fixed luminosity increasing from 10--30 per
cent of the population at the lowest densities, to $\sim 70$ per cent
of the population in the highest density
environments.  This is stronger than the dependence on luminosity at
fixed density; in particular, the trend with density is of a similar magnitude at all
luminosities.
The fraction of red--distribution galaxies (at
fixed luminosity) within the virial radius of clusters is 
independent of velocity dispersion, within the fairly large uncertainties.
This implies that the population differences are primarily related to
local galaxy density and not cluster mass or dynamics, in agreement with
other work \citep{Dressler,F+01,deP_BO}.

\begin{figure}
    \leavevmode\epsfxsize=8cm\epsfbox{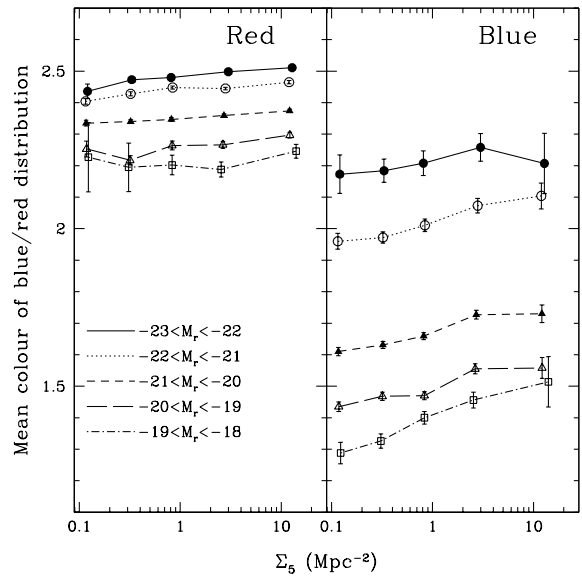}
    \caption{The mean color of the red distribution ({\it left panel})
      and the blue distribution ({\it right panel})
      as a function of
      local density, based on the Gaussian fitting of
      Fig.~\ref{fig-histograms}.  Five lines are shown, corresponding
      to five  different luminosity ranges, as indicated.  Points with
      $1-\sigma$ error bars are shown at the median density of each bin.
\label{fig-means}}
\end{figure}

In contrast to the strong trend in the fraction of red galaxies with $\Sigma_5$, the mean
color of each distribution depends only weakly on environment, as shown in
Figure~\ref{fig-means}.  Over a factor $\sim100$ in density, the
mean color of the blue (red) population changes by only $0.1$--$0.14$
($0.03$--$0.06$) 
mag, depending on luminosity; the strongest
trend is in the faint, blue galaxy distribution.  These trends are
weak, relative to the luminosity dependence: for example, the mean color of the blue population is $\sim 0.7$
mag redder in the brightest galaxies, compared with the faintest
galaxies in the same environment.  Similarly, we see little evidence that the dispersion
depends on environment: the reduced $\chi^2$ of the fits do not improve
significantly if the dispersions are allowed to vary,
and there is no significant trend in the best-fit dispersions with
environment.

\section{Discussion and Conclusions}\label{sec-discuss}
It has been known for a long time that the galaxy color distribution
(and correlated quantities)
depends on both environment \citep[e.g.][]{MS77,BO78b,Dressler,MS81} and
luminosity \citep[e.g.][]{dV61,BLE}.  We have shown here that the
two effects are separable if the full color distribution is considered as
the sum of two populations: while the mean and dispersion in the color of each 
type depend strongly on luminosity (Paper~I), they are weak
functions of environment.  For the red galaxy distribution, this is
well known \citep[e.g.][]{SV78}, and is consistent with the
idea that these are old galaxies, for which color is only weakly
sensitive to the present age \citep{Andreon03}.  

However, it is a surprising result that the colors of
blue galaxies, which are still actively growing and evolving, 
show such little dependence on environment, a result that was
indirectly revealed through earlier analysis of the Tully--Fisher relation \citep[e.g.][]{GMM}.
For example, if the increased abundance of red galaxies is due to interactions
(e.g. mergers and harassment) that increase in frequency
monotonically and smoothly with local galaxy density, we would expect
the blue distribution to become increasingly non-Gaussian with density,
and to gradually blend into the red distribution.
It seems unlikely that the small change of $\sim 0.1$ magnitudes in the
mean color of the blue distribution as a function of environment can be related to the much
larger change in the abundance of this population, relative to the red
population.  Instead, it may indicate a small difference in the
recent star formation history, or the distribution of intrinsic properties (e.g. dynamical mass 
or stellar velocity dispersion)  of blue galaxies at fixed luminosity
in different environments.

We propose that characteristic properties (e.g. metallicity,
dust content, and past-averaged SFR) of the late-type galaxies
are determined primarily by their luminosity (likely through its
relation to mass or other fundamental, intrinsic quantities) and that only
interactions of a certain level trigger a transformation
from late to early type.  This transformation must be either
sufficiently rapid, or sufficiently rare, to keep the overall color
distribution unchanged.  Furthermore, the mechanism responsible for
this transition must be effective for both bright and faint galaxies,
as the trends with local density are of a similar magnitude for both populations.

To quantify how rapid this transformation may be, we use
the \citet{BC03} stellar population
models to follow the color evolution from blue to red, assuming a
range of exponential transformation timescales $\tau$.  We start with a solar-metallicity
model of a 9 Gyr old galaxy with a \citet{Sp} initial mass function,
constant SFR, and a two component dust model with an effective extinction
of 1 mag, of which 0.3 mag arises from the ambient interstellar medium \citet{CF00}.
\citet{Brinchmann03} have shown that this star formation history is a
good description of the local, star forming population.
We then assume the red population has been built up by a steady
truncation of star formation in blue galaxies, over a Hubble time.  If
the truncation is rapid ($\tau<0.5$ Gyr), the color reddens to
$(u-r)_\circ\sim2.3$ in only 0.75 Gyr.  This change is so rapid that
we expect only $\sim 1$ per cent of the total galaxy population to be observed in the
mid-transition phase, defined as the $\sim 0.2$ magnitude dip
observed between the two Gaussian distributions.  From
Figure~\ref{fig-histograms}, it is evident that this would amount to an
increase of only $\sim 10$ per cent in these intermediate-color bins,
comparable to the observational uncertainties.
Therefore, the observed color distribution is 
not significantly altered by the presence of this transforming population, and the
simple Gaussian model remains a good fit to the data.  

On the other hand, if the SFR in a blue galaxy decays with
with an exponential timescale of 2 Gyr, it takes $\sim 4$ Gyr for the
colors to become as red as the early--type population.  In this case,
the number of galaxies with intermediate colors at the present day would
increase by a factor $\sim 2$, and would distort the
observed distribution in a way that is inconsistent with the
observations.  However, we caution that these results are sensitive to
the assumption that the transformations occur uniformly in time.  If,
instead, transitions were more common in the past, the data can accomodate a
slower rate of SFR decay, as expected if galaxies are stripped of their
hot gas in dense environments \citep{LTC,infall}.

We therefore conclude that short-timescale transformations could play a
role at all densities and luminosities, without disrupting the
Gaussian model fit. 
The small fraction of galaxies predicted to be in the transition phase 
is comparable to the fraction of spectroscopically
identified post-starburst galaxies \citep{tomo-EA1,Q+04_short}, and
anemic (passively evolving) spirals \citep{vdB76,Goto_passive_short} which
may be the signature of such transformations \citep[e.g.][]{DG92}.   

It is interesting to compare these results with the
morphology--density relation \citep{Dressler,PG84}.  In contrast with
the bimodality of the color
distribution, there are at least three morphological types
that have different dependences on environment: ellipticals, S0s, and
spirals.   One possibility is that the transforming
galaxies correspond to the S0 population, since morphology may change
on a longer timescale than color if the change is due to a decline in
SFR \citep{Bekki02}.  This interpretation is in qualitative
agreement with the dearth of such galaxies at higher redshift
\citep{D+97_short}.  More detailed analysis will be the subject of future work,
as morphologies for the SDSS galaxies become available \citep[e.g.][]{KM03}.

Finally, we note that, at all magnitudes, a population of red galaxies exists even in
low--density environments; this is consistent with Paper~II,
where we found a population of galaxies without significant H$\alpha$
emission in all environments.  Therefore, either some fraction of the
red population must arise independently of environment (e.g., by
consumption of the internal gas supply), or these are fossil groups 
which result from the complete merging of bright
galaxies  \citep{Ponman_fossil,MZ99,SHARC_catshort}.

\acknowledgements
We acknowledge support from PPARC fellowships
PPA/P/S/2001/00298 (MLB), PPA/Y/S/2001/00407 (RGB), and the David and
Lucille Packard foundation (IKB, KG).
\bibliography{ms}

\end{document}